\documentclass[twocolumn,amsmath,amssymb,floatfix,nofootinbib]{revtex4}
\usepackage{epsfig}
\usepackage{amsmath,amssymb}
\usepackage[all]{xy}

\usepackage{graphicx}
\usepackage{epstopdf}
\newcommand{\lsi}{\,\raisebox{-0.13cm}{$\stackrel{\textstyle<}
{\textstyle\sim}$}\,}
\newcommand{\gsi}{\,\raisebox{-0.13cm}{$\stackrel{\textstyle>}
{\textstyle\sim}$}\,}
\newcommand{\beq}{\begin{equation}}
\newcommand{\eeq}{\end{equation}}

\begin{document}

\title{Dark Matter and the Baryon Asymmetry}

\author{Glennys R. Farrar and Gabrijela Zaharijas}

\affiliation{Center for Cosmology and Particle Physics,
Department of Physics\\ New York University, NY, NY 10003,USA\\
}

\begin{abstract}{We present a mechanism to generate the baryon asymmetry of the Universe which preserves the net baryon number created in the Big Bang.  If dark matter particles carry baryon number $B_X$, and  
$\sigma^{\rm annih}_{\bar{X}} < \sigma^{\rm annih}_{X } $, the $\bar{ X}$'s freeze out at a higher temperature and
have a larger relic density than $X$'s.   If $m_X \lsi  4.5 \,B_X \,$GeV and the annihilation cross sections differ by $\mathcal{O}$(10\%) or more,  this type of scenario naturally explains the observed $\Omega_{DM} \approx 5\, \Omega_b$.   Two concrete examples are given, one of which can be excluded on observational grounds.
}
\end{abstract}
 \maketitle

The abundance of baryons and dark matter (DM) in our Universe poses
several challenging puzzles:\\
$\bullet$ Why is there a non-zero net nucleon density and what
determines its value?\\ 
$\bullet$ What does dark matter consist of? \\
$\bullet$  Is it an accident that the dark matter density is roughly comparable
to the nucleon density, $\rho_{DM} = 5 ~\rho_N$? 

As pointed out by Sakharov\cite{sakharov}, baryogensis requires three conditions: non-conservation of baryon number,  violation of C and CP, and a departure from thermal equilibrium.  The last is provided by the expansion of the Universe and the first two are naturally present in unified theories and even in the standard model at temperatures above the electroweak  phase transition.  However in most approaches the origins of DM and the Baryon Asymmetry of the Universe (BAU) are completely unrelated and their densities could naturally differ by many orders of magnitude.  In this Letter we propose a new type of scenario, in which the observed baryon asymmetry is due to the {\it separation} of baryon number between ordinary matter and dark matter and not to a net change in the total baryon number since the Big Bang. (See \cite{other} for other papers with this aim.)  Thus the abundances of nucleons and dark matter are related.  The first Sakharov condition is not required, while the last two remain essential.  We give explicit examples in which  anti-baryon number is sequestered at temperatures of order 100 MeV.  There is no need for the Universe to reach significantly higher temperatures, so the number of e-foldings required in inflation is reduced, potentially ameliorating difficulties with Early Universe models.

CPT requires that the total interaction rate of any ensemble of particles and antiparticles is the same as for the  conjugate state in which each particle is replaced by its antiparticle and all spins are reversed.  However individual channels need not have the same rate so, when CP is violated, the annihilation rates of the CP reversed systems are not in general equal.  A difference in the annihilation cross section, $\sigma^{\rm annih}_{\bar{X}} <  \sigma^{\rm annih}_{X } $, means that the freeze out temperature for $X$'s ($T_X$) is lower than for $\bar{X}$'s ($T_{\bar{X}}$).   After the $\bar{X}$'s freeze out, the $X$'s continue to annihilate until the temperature drops to $T_{X}$, removing $B_X$ antinucleons for each $X$ which annihilates.  

Assuming there are no other significant contributions to the DM density, the present values $n_{o\, N}$, $n_{o\, X}$ and $n_{o\, \bar{X}}$  are
determined in terms of $m_X$, $B_X$ and the observables $ \frac{\Omega_{DM}}{\Omega_b}$ and $\frac{n_{o\,N}}{n_{o\, \gamma}} \equiv \eta_{10} \,10^{-10}$ or $\rho_{\rm crit}$.  (Following common usage, the subscript $b$ refers to the baryon number in ordinary matter.)  Observations yield $\eta_{10} = 6.5^{+0.4}_{-0.3}$, $\Omega_m h^2 = 0.14 \pm 0.02$, $\Omega_b h^2 = 0.024 \pm 0.001$\cite{WMAP},  so $\frac{\Omega_{DM}}{\Omega_b} = 4.83 \pm 0.87$.  Given the values of these observables, we can ``reverse engineer" the process of baryon-number-segregation.  

For brevity, suppose there is only one significant species of DM particle.  Let us define $\epsilon \equiv  \frac{n_X}{n_{\bar{X}} }$.  Then the total energy density in $X$'s and $\bar{X}$'s is  $\rho_{DM} = m_X n_{\bar{X}} (1 + \epsilon)$.  By hypothesis, the baryon number density in nucleons equals the antibaryon number density in $X $ and $\bar{X}$'s, so  $ B_X n_{\bar{X}} (1-\epsilon) = (n_N - n_{\bar{N}}) = \frac{\rho_b}{m_N}$.  Thus  
\begin{equation} \label{kappa}
\frac{\Omega_{DM}}{\Omega_b} = \left( \frac{1 + \epsilon}{1 - \epsilon} \right) \frac{m_X}{m_N B_X}.
\end{equation}
As long as the DM particle mass is of order hadronic masses and $\epsilon $ is not too close to 1, this type of scenario naturally accounts for the fact that the DM and ordinary matter densities are of the same order of magnitude.  Furthermore, since $ \frac{1 + \epsilon}{1 - \epsilon} \ge 1$, the DM density in this scenario must be {\it  greater} than the nucleonic density (unless $m_X < m_N B_X$), as observed.  

Given the parameters of our Universe, we can instead write (\ref{kappa}) as an equation for the DM mass
$
 m_X = \left( \frac{1 - \epsilon}{1 + \epsilon} \right) \frac{\Omega_{DM}}{\Omega_b} \, B_X m_N .
$
For low baryon number, $B_X = 1\, (2)$, this implies $m_X \lsi 4.5 \,(9)\,$GeV.  If dark matter has other components in
addition to the $X$ and $\bar{X}$, the $X$ must be lighter still.   The observed BAU can be due to baryon number sequestration with heavy DM only if $B_X$ is very large, e.g., strangelets or Q-balls.  However segregating the baryon number in such cases is challenging.  

As an existence proof and to focus our discussion of the issues, we present two concrete scenarios. In one, the DM particles are the $H$ (strangeness -2 dibaryon) and $\bar{H}$, with $m_X = 2$ GeV and $B_X = 2$;  CP violation beyond CKM is required in the Early Universe when $T\sim 100$ MeV.  In the second scenario we postulate a new particle $X$ with $B_X = 1$ and mass  $\lsi 4.5$ GeV, which couples to quarks through dimension-6 operators coming from beyond-the-standard-model physics; CP violation is naturally large enough, $\mathcal{O}$(10\%), because all three quark generations are involved and the new interactions in general also violate CP.  As we shall see, the $H,~\bar{H}$ model can be excluded in an almost model independent way, while the $B_X = 1$ model seems unassailable on experimental grounds, except that it will be excluded if DM-induced nuclear recoil is detected.

We begin by deducing the properties of the DM particle $X$ and their interactions, which are required in order to obtain the correct relic abundances of baryons and DM.  The annihilation rate of particles of type $j$ with particles of type $i$ is
$\Gamma^{annih}_{j}(T) = \Sigma_i ~n_i(T) \langle \sigma_{ij}^{annih} v_{ij}\rangle $, 
where $n_i$ is the number density of the $i$th species,  $v_{ij}$ is the relative velocity, and $\langle...\rangle$ indicates a thermal average.  As the Universe cools, the densities of all the particle species decrease and eventually the rate of even the most important
annihilation reaction falls below the expansion rate of the Universe.  The temperature at which this occurs is called the
freezeout temperature $T_j$, satisfying $\Gamma^{annih}_{j}(T_{j}) = H(T_{j}) = 1.66 \sqrt{g_*} ~ T_{j}^2/ M_{Pl} $, where $g_*$ is the effective number of relativistic degrees of freedom at temperature $T_{j}$ \cite{kolbTurner}.  Between a few MeV and the QCD phase transition the only relativistic degrees of freedom in equilibrium  are neutrinos, $e^\pm$ and $\gamma$'s and $g_* = 10.75$.  Above the QCD phase transition, which is estimated to be within the range 100 to 200 MeV, light quarks and antiquarks ($q, \, \bar{q}$) and $\mu^\pm$ are also relativistic species in equilibrium, giving $g_* = 56.25$.  As we shall see, freezeout is below the QCD phase transition in the $H,~\bar{H}$ scenario, and above it in our $B_X=1$ example according to standard estimates of $T_{QCD}$.   Relic abundances are in general determined by solving coupled Boltzmann equations, but to a good approximation are given by the equilibrium densities at freeze out temperature, $n_{\bar{X}}(T_{\bar{X}})$ and  $n_{{X}}(T_{{X}})$\cite{kolbTurner}.  

Given $m_X,\, B_X$ and $g_X$ (the number of degrees of freedom of the $X$ particle), the temperature $T_{\bar{X}}$ at which $\bar{X}$'s must freeze out to give the observed abundances satisfies: 
\begin{equation} \label{Xbarfo}
  \frac{n_{\bar{X}} - n_X}{n_{\bar{X}}} \frac{n_{\bar{X}}}{n_\gamma } = (1-\epsilon) \frac{ \pi^2 g_X
  x_{\bar{X}}^{3/2} e^{-x_{\bar{X}} } }{2 \zeta(3) (2 \pi)^{3/2} }= 
  \frac{10.75}{3.91}\frac{\eta_{10} 10^{-10} }{B_X} ,
\end{equation}
where $x_{\bar{X}} \equiv m_X/T_{\bar{X}}$.  
$ \frac{10.75}{3.91}$ is the factor by which $\frac{n_b}{n_\gamma}$ increases above $e^\pm$ annihilation.  The equation for $X$ freezeout is the same, with $(1-\epsilon) \rightarrow (1-\epsilon)/\epsilon $.  
Freezeout parameters for the two specific models are given in Table I; $\tilde{\sigma} \equiv \langle \sigma^{\rm ann} |v| \rangle / \langle |v| \rangle$ denotes the effective annihilation cross section, averaged over the relevant distribution of c.m. kinetic energies at freezeout.  

A key element of baryon-number sequestration is that self-annihilation cannot be important for maintaining equilibrium prior to freeze out, or else an asymmetry between $X$ and $\bar{X}$ cannot arise.  Due to their higher mass, $X,\bar{X}$ abundances are much lower than light quarks and nucleons at freeze-out.  Thus the required dominance of co-annihilation obtains as long as the co-annihilation cross section at freezeout in the most important channel for maintaining equilibrium, $\sigma_{\bar{X}}^{\rm co-ann}$,  is not too small -- specifically,  $ \sigma_{\bar{X} q}^{\rm ann}\gsi 10^{-11} \sigma^{\rm ann}_{\bar{X} X}$ and $ \sigma_{\bar{X} N}^{\rm ann}\gsi 10^{-5} \sigma^{\rm ann}_{\bar{X} X}$ in the $B_X = 1$ and $H,~\bar{H}$ scenarios respectively. 

\begin{table}[htb] \label{table}
\caption{Required freezeout temperatures and annihilation cross sections at freezeout.}
\begin{center}
\begin{tabular}{|c|c|c|c|c|}
\hline
Model & $T_{\bar{X}}$ MeV & $T_{X}$ MeV   &  $\tilde{\sigma}^{\rm ann}_{\bar{X}}$ cm$^2$ & $\tilde{\sigma}^{\rm ann}_{{X}}$ cm$^2$    \\  \hline 
$H,~\bar{H}$   &      86.3         &   84.5    & 	$2.2~10^{-41}$                       &    $2.8~10^{-41}$                   \\ \hline
$B_X = 1$   &      180          &   159     &		$1.3~10^{-43}$                       &    $5.7~10^{-43}$                     \\ \hline
\end{tabular}
\end{center}
\end{table}

CPT requires that $\sigma^{\rm ann}_{X} + \sigma^{\rm non-ann}_{X} = \sigma^{\rm ann}_{\bar{X}} + \sigma^{\rm non-ann}_{\bar{X}}$.  Therefore a non-trivial consistency condition in this scenario is 
$
\sigma^{\rm ann}_{X} - \sigma^{\rm ann}_{\bar{X}} \le \sigma^{\rm non-ann}_{\bar{X}}.
$
The value of the LHS needed for B-sequestration (see Table I) is comfortably compatible with $\sigma^{\rm non-ann}_{\bar{X}} \ge \sigma^{\rm el}_{\bar{X}} $ and the upper limits on $\sigma^{\rm non-ann}_{\bar{X}}$ from DM searches.

Nucleosynthesis works the same way in these scenarios as for standard CDM, because $\sigma_{\{X,\bar{X} \}N}$ is negligible compared to $\sigma_{NN}$ and the $X,\,\bar{X}$ do not bind to nuclei\cite{fz:nucbind}.  Thus primordial light element abundances actually constrain the {\it nucleon} -- rather than {\it baryon} -- to photon ratio. 

We now turn to whether it is possible to achieve the required particle physics in the $B_X = 1,~2$ scenarios, and then to observational constraints.  The essential particle physics requirements are that the lifetime of the DM particle be larger than the age of the universe, the properties of the DM particle be consistent with accelerator constraints, the scattering cross section of $X$ and $\bar{X}$ from nucleons be consistent with present limits from DM detectors, and that annihilation of $\bar{X}$ with nucleons in Earth, Sun, galactic center and elsewhere not be observed.   Remarkably, if the H is compact, $r_H \lsi r_N/(2-3)$, and lighter than $m_\Lambda + m_N$, its lifetime could be sufficiently long and it would not be excluded by double-$\Lambda$ hypernuclei experiments, seen in H production experiments, or lead to unobserved exotic isotopes \cite{fz:nucbind}, so from a purely particle physics standpoint the $H,\, \bar{H}$ DM scenario is viable at present; details and additional references are given in \cite{fzBAU04}.
Now we show that our other scenario, a light fundamental particle with $B_X = 1$,  $m_X \lsi 4.5$ GeV and the required coupling to quarks is also consistent with laboratory and observational constraints.  

In order to be sufficiently weakly coupled (see Table I), the $X$ in the $B_X = 1$,  $m_X \lsi 4.5$ GeV scenario should be a standard-model singlet, and its low mass suggests it may be a fermion whose mass is protected by a chiral symmetry.  Then, dimension-6 interactions with quarks could appear in the low energy effective theory after high scale interactions, e.g., those responsible for the family structure of the Standard Model, have been integrated out, for instance
\begin{equation} \label{Xbcd}
\kappa (\bar{X} b \, \bar{d^c} c - \bar{X} c \, \bar{d^c} b) + h.c.,
\end{equation}
where the
$b$ and $c$ fields are left-handed SU(2) doublets combined to form an SU(2) singlet and $d^c$ is the charge conjugate of the SU(2) singlet field $d_R$, and $\kappa \sim g^2/\Lambda ^2$, where $\Lambda$ is an energy scale of new physics.  The suppressed color and spin indices are those of the antisymmetric operator $\tilde{O}^{\dot{a}}$ given in equation (10) of ref. \cite{peskin79}.  The $X$ is a singlet under all standard model interactions and its only interaction with fields of the standard model are through operators such as (\ref{Xbcd}).  Note that $\kappa$ is in general temperature dependent; we denote its value today and at freezeout by $\kappa_0$ and $\kappa_{\rm fo}$.  Prior to freezeout, $\bar{X}$'s stay in equilibrium through reactions like 
\begin{equation}
\label{dXbar}
d + \bar{X} \leftrightarrow \bar{b}~\bar{c}. 
\end{equation}
The coupling $\kappa $ in (\ref{Xbcd}) is in general complex and a variety of diagrams involving all three generations and including both W exchange and penguins contribute to generating the effective interaction in (\ref{dXbar}), so the conditions necessary for a sizable CP-violating asymmetry between $\sigma_{X}^{ \rm ann} $ and $ \sigma_{\bar{X}}^{ \rm ann}$ are in place.  An interaction such as (\ref{Xbcd}) leads to an annihilation cross section at freezeout
\begin{equation} 
\sigma_{ \bar{b} \bar{c}\rightarrow \bar{X} d} \approx \kappa^2_{\rm{fo}} \sqrt{\frac{m_c}{T_{\rm fo} }} \frac{m_X [ (m_b + m_c)^2 - m_X^2]^2}{(m_b+m_c)^3}.
\end{equation} 
For the freezeout of $X$ to occur at the correct temperature (see Table 1), $\kappa_{\rm{fo}} \approx 10^{-8}\, {\rm GeV}^{-2}$ is needed. This suggests an energy scale for new physics of $\Lambda \lsi 10$ TeV, taking dimensionless couplings to be  $\lsi 1$. 

$X$ particles can decay via production of $bcd$ quarks. For an $X$ particle mass of 4.5 GeV the decay is off-shell, with a W exchange between b and c quarks, giving
$
X\rightarrow csd.
$
The decay rate of $X$ at zero temperature can be estimated as:
\begin{equation}
\Gamma \sim m^5 _X \kappa ^2 _0 g^4 _W |V_{bc}V_{cs}|^2,
\end{equation}
where $g_W$ is the electroweak $SU(2)$ gauge coupling and the $V$'s are mixing angles.  The condition $\tau _X \gsi \tau_{\rm Univ}$ constrains the $X$ coupling today: $\kappa _{0} \lsi 10^{-20}$ GeV$^{-2}$.  Thus for $X$ to be a valid dark matter candidate, its coupling to ordinary matter needs to have a strong temperature dependence, changing from $10^{-8}\, {\rm GeV}^{-2}$ at a temperature of $\sim 180$ MeV, to $10^{-20}\, {\rm GeV}^{-2}$ or effectively zero, at zero temperature.  The most attractive way to do this would be if the interaction (\ref{Xbcd}) were related somehow to a sphaleron-type phenomenon which was allowed above the QCD or chiral phase transition, but strongly suppressed at low temperature.  We do not attempt to develop such a model here, but instead give two examples that show the desired dramatic change in $\kappa$ is possible.  

If the interaction (\ref{Xbcd}) is mediated by a field $\eta$ with couplings of order 1, its effective mass $m_{\eta}$ should vary from 10 TeV at $T \approx 180$ MeV to $10^{10}$ TeV at zero temperature. Let the $\eta$ mass be due to the VEV of a neutral scalar field $\sigma$ and a bare mass term.  The VEV of $\sigma$ can be rapidly driven from zero to some fixed value resulting in the desired mass change in $\eta$ by several possible mechanisms.  The simplest requires only one additional field with the zero temperature interaction
\begin{equation}V(\eta ,\sigma )=-m^2 _{\sigma}\sigma ^2+\alpha_1 \sigma ^4 +\alpha_2 \eta ^4 +2\alpha_3 \sigma ^2 \eta ^2.
\end{equation}
The global minimum of this potential at zero temperature is at $\langle \eta \rangle =0,~ \langle \sigma ^2\rangle = \pm m^2 _{\sigma}/(2\alpha _1)  $, giving
$ m_{\eta ,0}=\sqrt{2\alpha_3 \sigma ^2 }=\sqrt{(\alpha_3/\alpha_1) } m_{\sigma}$, which must be $\gsi 10^{10}$ TeV to assure sufficiently long-lived DM.  At higher temperature, one loop corrections contribute to the potential and introduce a temperature dependence \cite{weinberg74,senjanovic}:
\begin{equation}
V_{\rm {loop}}=\frac{2\alpha_1+\alpha_3}{6}T^2\sigma ^2+\frac{2\alpha_2+\alpha_3}{6}T^2\eta ^2.
\end{equation}
The new condition for the minimum of the potential in the $\sigma$ direction becomes
\begin{equation}
\langle \sigma ^2\rangle ~=\frac{m^2 _{\sigma}-T^2\left( 2\alpha _1+\alpha _3\right)/3}{4\alpha _1},
\end{equation} 
and for temperatures higher than the critical value 
$T_{CR}=\frac{3m^2 _{\sigma}}{2\alpha _1+\alpha _3}$
the potential has a unique minimum at $\langle \eta\rangle =0$, $\langle \sigma \rangle =0$. To obtain $T_{CR}\sim 180$  MeV with the conditions above implies
$ \alpha _3\sim 10^7\sqrt{\alpha _1}.  $ This looks fine-tuned, but that defect can be removed by introducing another scalar field $\phi$ as in  Linde's hybrid inflation models \cite{linde}.  In this scenario a very light overall-singlet scalar is required, which presents naturalness challenges but seems otherwise to be consistent with cosmology; details can be found in ref. \cite{z:thesis}.

The approach presented here to solve the DM and BAU puzzles at the same time, with anti-baryonic dark matter, can run afoul of observations in several ways which we now check.  The scattering cross sections $\sigma_{XN}$ and $\sigma_{\bar{X}N}$ must be compatible with DM searches.  The H, remarkably, passes this test at present because for $\sigma_{HN}$ and $\sigma_{\bar{H}N}$ of order $\mu$b -- a reasonable cross section in the compact-H scenario \cite{f:StableH} -- there is a window in the DM limits for $M_{DM} \lsi 2.2$ GeV \cite{zf:window}. The allowed window could be easily closed, but it is fascinating that it falls in the right range for the $H$ DM model.  The $X$ and $\bar{X}$ are more like conventional DM so standard WIMP searches constrain the low energy scattering cross section  $\sigma_{DM} \equiv (\sigma^{\rm el}_{\bar{X} N} + \epsilon \sigma^{\rm el}_{XN})/(1+ \epsilon)$;  the present limit is $\sigma_{DM} \lsi 10^{-38} {\rm cm}^2$ for a 4 GeV particle.  It is not possible to predict $\sigma_{XN}$ and $\sigma_{\bar{X}N}$ for a given value of $\kappa_0$ from the $X$ lifetime without understanding how the interaction (\ref{Xbcd}) is generated, since it is not renormalizable.  A naive guess 
\begin{equation}
\sigma^{\rm el}_{XN \{\bar{X} N\} } \sim \kappa^4 \Lambda ^4 \frac {m^2 _X m^2 _N}{(m_X+m_N)^2}\sim 10^{-67}~{\rm cm}^2
\end{equation}
is far out of reach of currently imagined sensitivities using the maximum allowed value of $\kappa_0$, but the actual value of the scattering cross section on nucleons depends on the high-scale physics and could be significantly larger or smaller. 

Another consideration, not present in conventional DM scenarios, arises because of the possibility of annihilation of the anti-baryonic component of DM with nucleons.  Since DM does not concentrate to the extent that nucleons do, this can be much more important than self-annihilation in conventional WIMP scenarios.  The final products of $\bar{X} N$ and $\bar{H} N$ annihilation are mostly pions and some kaons, with energies of order 0.1 to 1 GeV.  If the DM scattering cross section on nucleons is large enough that a significant fraction $f_{cap}$ of the incident DM particles is captured in a planet or star of effective cross-sectional area $S$, and evaporation can be neglected as is the case here, we have the following model-independent lower bound on the power released in annihilation, independent of the annihilation cross section and the mass of the DM particle: 
\beq \label{power}
{\rm Power} \ge  f_{cap} S \langle \rho_{DM}  v_{DM} \rangle   ,
\eeq
where $S \langle \rho_{DM}/m_X~  v_{DM} \rangle$ is the captured number of DM particles per unit time.  Equation (\ref{power}) follows because every annihilation releases $m_X + B_X m_N \ge m_X$ of energy.  Uranus is large and has extremely low insolation, and there is no evidence of any internal source of heat\cite{uranusVoyager}.  With a $\mu$b scattering cross section and using the Edsjo et al flux calculation\cite{edsjo}, the predicted power in $\bar{H}$ annihilation in Uranus is $\approx 10$ times greater than the experimental upper limit, (the predicted power is $470$ erg cm$^2$s$^{-1}$, while the measured value is $42\pm 47$ erg cm$^2$s$^{-1}$) \cite{uranusVoyager}.  Thus we can exclude the $H,\bar{H}$ dark matter scenario without delving into the difficult particle physics dynamics of the $H$ which would determine, for instance, the $\bar{H}$ annihilation cross section. (Ref. \cite{GN} uses annihilation in the sun, along with guesses about $H$ and $\bar{ H}$ properties, to argue against the $H,~\bar{ H}$ DM scenario.)

However if the DM is a new particle with very weak elastic scattering interactions with nucleons, a negligible fraction of the DM flux is captured and (\ref{power}) does not apply.  In this case, the very stability of the $X$ implies that its interaction with light quarks is so weak that its annihilation rate in the $T=0$ universe is almost infinitesimally small. The cross section for the dominant annihilation processes such as $\bar{X} N \rightarrow \bar{D} K$ is governed by the same Feynman diagrams (in a crossed channel) as govern $X$ decay; dimensional arguments lead to the order-of-magnitude relation 
\beq 
\label{sigann}
\sigma^{ann}_{\bar{X}N} v \sim m_X^{-3} \tau_X^{-1}  \lsi 10^{-72} (30 {\rm Gyr}/\tau_X) {\rm cm}^2 .
\eeq  
DM in this model will be virtually impossible to detect directly or through its annihilations, unless the scattering cross section is much larger than the naive estimate (\ref{sigann}).  One might hope to find direct evidence for such an $X$ particle via the existence of strikingly exotic final states in rare B decay, e.g., $B \rightarrow X \bar{\Lambda_c}$ + mesons, which would have considerable missing energy and seemingly violate baryon number.  Unfortunately, the branching fraction for such a decay would be of order $(\kappa_0/G_F)^2 \lsi 10^{-30}$ rendering it effectively unobervable.  Similarly, production of $X$'s and $\bar{X}$ in accelerators even at LHC energies $\sim \Lambda$ would be -- at least naively -- negligibly small.

To summarize, we have shown that the dark matter and baryon asymmetry puzzles may be related.  We presented two concrete scenarios, in which the observed values of $\Omega_{DM}$ and $\Omega_b$ are explained.  One of them entails a long-lived H dibaryon, but we excluded that scenario by limits on heat production in Uranus.  The other case involves a new  $\sim 4$ GeV particle with dimension-6 couplings to quarks.  Although the required parameters for this scenario can be seem to be achievable with beyond-the-standard-model particle physics, our analysis shows that the required new DM particle will be virtually impossible to detect with DM detection experiments, accelerator experiments, or astrophysical searches for annihilation products.  Thus unless the model is excluded by discovery of a signal in DM detection experiments, it will have to be judged primarily on theoretical grounds.   \\

\noindent{\bf Acknowledgements: } 
This research was supported in part by NSF-PHY-0101738.  We have benefited from discussions with many colleagues; special thanks go to S. Mitra for bringing the strong limits on heating in Uranus to our attention, and to S. Nussinov for discussions about ways to exclude the H dibaryon scenario, in particular the idea of using white dwarf heating which although unsuccessful planted the idea we used here with Uranus. GRF also thanks the Departments of Physics and Astronomy of Princeton University for their support and hospitality during the course of this research.


\end{document}